\documentclass[12pt]{iopart}
\usepackage[dvips]{graphicx}
\usepackage{hyperref}
\usepackage{bm}
\usepackage{color}

\eqnobysec

\begin{document}


\title{Dynamical description of Tychonian Universe}

\author{Luka Popov \vspace{2mm}}

\address{University of Zagreb, Department of Physics,
Bijeni\v cka cesta 32, Zagreb, Croatia}

\ead{lpopov@phy.hr}

\begin{abstract} \noindent
Using Mach's principle, we will show that the observed diurnal and annual motion of
the Earth can just as well be accounted as the diurnal rotation and annual revolution
of the Universe around the fixed and centered Earth. This can be performed by postulating
the existence of vector and scalar potentials caused by the simultaneous motion of
the masses in the Universe, including the distant stars.
\end{abstract}





 \maketitle

\setcounter{equation}{0}
\section{Introduction}\label{intro}

The modern day use of the word \emph{relativity} in physics is usually connected
with Galilean and special relativity, i.e.~the equivalence of the systems performing
the uniform rectlinear motion, so-called \emph{inertial frames}. Nevertheless,
the physicists and philosophers never ceased to debate the various topics under
the heading of \emph{Mach's principle}, which essentially claims the equivalence
of all co-moving frames, including non-intertial frames as well.

Historically, this issue was first brought out by Sir Isaac Newton in his famous
rotating bucket argument. As Newton saw it, the bucket is rotating in the absolute
space and that motion produces the centrifugal forces manifested by the concave shape
of the surface of the water in the bucket. The motion of the water is therefore to
be considered as ``true and absolute'', clearly distinguished from the relative motion
of the water with respect to the vessel \cite{principia}.

Mach, on the other hand, called the concept of absolute space a ``monstrous conception''
\cite{mach1}, and claimed that the centrifugal force in the bucket is the result only
of the relative motion of the water with respect to the masses in the Universe.
Mach argued that if one could
rotate the whole Universe around the bucket, the centrifugal forces would be generated,
and the concave-shaped surface of the water in the bucket would be identical as in the
case of rotating bucket in the fixed Universe. Mach extended this principle to the
once famous debate between geocentrists and heliocenstrists, claiming that both systems
can equally be considered correct \cite{mach2}.

His arguments, however, remained of mostly philosophical nature. Since he was convinced
empiricist, he believed that science should be operating only with observable facts, and
the only thing we can observe are relative motions. Therefore, every notion of absolute
motion or a preferred inertial frame, whether inertial or non-inertial, is not a scientific
one but rather a mathematical or philosophical preference.

As Hartman and Nissim-Sabat \cite{hartman} correctly point out,
Mach never formulated the mathematical model or an alternative set of physical laws which
can explain the motions of the stars, the planets, the Sun and the Moon in a Tychonian
or Ptolemaic geocentric systems. For that reason, some physicist in the modern days have
tried to ``Machianize'' the Newtonian mechanics in various ways \cite{hood, barbourN},
or even try to construct new theories of mechanics \cite{assis}. There have also been
attempts to reconcile Mach's principle with the General Theory of Relativity, some of
which were profoundly analyzed in the paper by Raine \cite{raine}.

In the recent paper \cite{popov1} we have used the concept of the so-called
pseudo-force and derived the expression for the potential which is responsible for it.
This potential can be considered as a real potential (as shown by Zylbersztajn
\cite{zylbersztajn}), which can easily explain the annual motion of the Sun and planets
in the Neo-Tychonian system. In the same manner, one can explain the annual motion of the
stars and the observation of the stellar parallax \cite{popov2}.

It is the aim of this paper to use the same approach to give the dynamical explanation of the
diurnal motion of the celestial bodies as seen from the Earth, and thus give the mathematical
justification for the validity of Mach's arguments regarding the equivalence of the Copernican
and geocentric systems.

The paper is organized as follows. In section \label{vectorpot} the vector potential is
introduces in general terms. This formalism is then applied to analyze the motions of
the celestial bodies as seen from the Earth in section \label{celestmot}. Finally, the
conclusion of the analysis is given.

\section{Vector potential formalism}\label{vectorpot}

Following Mach's line of thought, one can say that the simultaneously rotating Universe
generates some kind of gravito-magnetic vector potential, $\mathbf{A}$. By the analogy
with the classical theory of fields \cite{classf} one can write down the
Lagrangian which includes the vector potential,
\begin{equation}\label{Lgen}
L = \frac{1}{2} m \dot{\mathbf{r}}^2 + m\, \dot{\mathbf{r}} \!\cdot\! \mathbf{A} +
    \frac{1}{2} m \mathbf{A}^2 - m U_{\mathrm{ext}}\,,
\end{equation}
where $m$ is the mass of the particle under consideration, and
$U_{\mathrm{ext}}$ is some external scalar potential imposed on the particle, for
example the gravitational interaction.

We know, as an observed fact, that every body in the rotational frame of reference
undergoes the equations of motion given by \cite{goldstein}
\begin{equation}\label{eom}
m \ddot{\mathbf{r}} = \mathbf{F}_{\mathrm{ext}} - 2 m (\bm\omega_{\mathrm{rel}} \times \dot{\mathbf{r}}) -
    m \left[ \bm\omega_{\mathrm{rel}} \times (\bm\omega_{\mathrm{rel}} \times \mathbf{r}) \right] \,,
\end{equation}
where $\bm\omega_{\mathrm{rel}}$ is relative angular velocity between the given frame of
reference and the distant masses in the Universe, and
$\mathbf{F}_{\mathrm{ext}} = -\nabla U_{\mathrm{ext}}$ some external force acting on a particle.

It can be easily demonstrated that one can derive Equation (\ref{eom}) by applying the
Euler-Lagrange equations on the following ``observed'' Langrangian
\begin{equation}\label{Lobs}
L_\mathrm{obs} = \frac{1}{2} m \dot{\mathbf{r}}^2 +
    m\, \dot{\mathbf{r}} \cdot (\bm\omega_{\mathrm{rel}} \times \mathbf{r}) +
    \frac{1}{2} m (\bm\omega_{\mathrm{rel}} \times \mathbf{r})^2 -
    m U_{\mathrm{ext}}\,.
\end{equation}
By comparison of the general Lagrangian (\ref{Lgen}) and the ``observed'' Lagrangian
(\ref{Lobs}) one can write down the expression for the vector potential $\mathbf{A}$,
\begin{equation}\label{vecpot}
\mathbf{A} = \bm\omega_{\mathrm{rel}} \times \mathbf{r} \,.
\end{equation}

It is important to notice that there is no notion of the absolute rotation in this
formalism. The observer sitting on the edge of the Newton's rotating bucket can only
observe and measure the relative angular velocity between him or her and the distant stars
$\bm\omega_{\mathrm{rel}}$, incapable of determine whether it is the bucket
or the stars that is rotating.

\section{Trajectories of the celestial bodies around the fixed Earth}\label{celestmot}

\subsection{Diurnal motion}

It is one thing to postulate that rotating masses in the Universe generate the
vector potential given by (\ref{vecpot}), but quite another to claim that this
same potential can be used to explain and understand the very motion of these
distant masses. We will now demonstrate that this is indeed the case.

The observer sitting on the surface of the Earth makes several observations. First,
he or she notices that there is a preferred axes (say $z$) around which all Universe
rotates with the period of approximately 24 h. Then, according to the formalism given in Section
\ref{vectorpot}, he or she concludes that the Earth must be immersed in the vector potential
given by
\begin{equation}\label{Evecpot}
\mathbf{A} =  \Omega \; \hat{\mathbf{z}} \times \mathbf{r} \,,
\end{equation}
where $\Omega \approx (2\pi / 24 \textrm{ h})$ is the observed angular velocity of the celestial
bodies \footnote{The period of the relative rotation between the Earth and the distant stars is
called \emph{sidreal day} and it equals 23 h 56' 4.0916''. Common time on a typical clock measures
a slightly longer cycle, accounting not only for the Sun's diurnal rotation but also for the
Sun's annual revolution around the Earth (as seen from the geocentric perspective) of slightly less
than 1 degree per day \cite{wiki}.}.

One can now re-write the Lagrangian (\ref{Lgen}) together with the Equation
(\ref{Evecpot}) and focus only on the contributions coming from the vector potential
 $\mathbf{A}$,
\begin{equation}\label{lang2}
L_\mathrm{rot} = \frac{1}{2} m \dot{\mathbf{r}}^2 +
    m\, \Omega\, \dot{\mathbf{r}} \cdot (\hat\mathbf{z} \times \mathbf{r}) +
    \frac{1}{2} m\, \Omega^2\, (\hat\mathbf{z} \times \mathbf{r})^2 \,.
\end{equation}

The Euler-Lagrange equations for this Lagrangian, written for each component of the
Cartesian coordinates, are given by
\begin{eqnarray}\label{eomsCart}
\ddot{x} & = & -2 \, \Omega \, \dot{y} + \Omega^2 \, x \nonumber \\
\ddot{y} & = & 2 \, \Omega \, \dot{x} + \Omega^2 \, y  \\
\ddot{z} & = & 0 \nonumber \,.
\end{eqnarray}
The solution of this system of differential equations reads
\begin{eqnarray}\label{sol1}
x(t) & = & r \, \cos \Omega t \nonumber \\
y(t) & = & r \, \sin \Omega t \\
z(t) & = & 0 \nonumber \,,
\end{eqnarray}
where $r$ is the initial distance of the star from the $z$ axes. The observer can
therefore conclude that the celestial bodies perform real circular orbits around the static
Earth due to the existence of the vector potential $\mathbf{A}$ given by
Equation (\ref{Evecpot}). This conclusion is equivalent to the one that claims that the
Earth rotates around the $z$ axes and the celestial bodies don't.

\subsection{Annual motion}

The second thing the observer on the Earth notices is the periodical annual motion
of the celestial bodies around the $z'$ axes which is inclined
form the axes of diurnal rotation $z$ by the angle of approximately $23.5^\circ$.
This motion can be explained if one assumes that the Earth is immersed in the
so-called pseudo-potential
\begin{equation} \label{Ups}
U_{\mathrm{ps}} (\mathbf{r}) = \frac{G M_S}{r_{SE}^2} \hat{\mathbf{r}}_{SE} \cdot
\mathbf{r} \,.
\end{equation}
Here $G$ stands for Newton's constant, $M_S$ stands for the mass of the Sun and
$\mathbf{r}_{SE}(t)$ describes the motion of the Sun as seen form the Earth. The Sun's
trajectory  $\mathbf{r}_{SE}(t)$ is shown to be an ellipse in $x'$-$y'$ plane
(defined by the $z'$ axes from the above). Using this potential alone
one can reproduce the observed retrograde motion of the Mars or explain
the effect of the stellar parallax as the real motion of the distant stars in
the $x'$-$y'$ plane. All this was demonstrated in the previous communications
\cite{popov1,popov2}.

\subsection{Total account}

One can finally conclude that all celestial bodies in the Universe perform the
twofold motion around the Earth:
\begin{enumerate}
\item circular motion in the $x$-$y$ plane due to the vector potential
    $\mathbf{A}$ (\ref{Evecpot}) with the period of approximately 24 hours and
\item elliptical orbital motion in the $x'$-$y'$ plane due to the scalar potential
    $U_{\mathrm{ps}}$ (\ref{Ups}) with the period of approximately one year.
\end{enumerate}

Using Equations (\ref{Lgen}), (\ref{Evecpot}) and (\ref{Ups}) one can write down
the complete classical Lagrangian of the geocentric Universe,
\begin{eqnarray}\label{Lcompl}
L & = & \frac{1}{2} m \dot{\mathbf{r}}^2 +
    m\, \Omega\, \dot{\mathbf{r}} \cdot (\hat\mathbf{z} \times \mathbf{r}) +
    \frac{1}{2} m\, \Omega^2\, (\hat\mathbf{z} \times \mathbf{r})^2 \nonumber \\
 & & {}- m
    \frac{G M_S}{r_{SE}^2} \hat{\mathbf{r}}_{SE} \cdot \mathbf{r} -
    m U_{\mathrm{loc}} \,,
\end{eqnarray}
where $U_{\mathrm{loc}}$ describes some local interaction, e.g.~between the
planet and its moon.

It is a matter of trivial exercise to show that these potentials can easily
account for the popular ``proofs'' of Earth's rotation like the Faucault's
pendulum or the existence of the geostationary orbits.

\section{Conclusion}

We have presented the mathematical formalism which can justify Mach's statement
that both geocentric and Copernican modes of view are ``equally actual'' and
``equally correct'' \cite{mach2}. This is performed by introducing two potentials,
(1) vector potential that accounts for the diurnal rotations and (2) scalar potential that
accounts for the annual revolutions of the celestial bodies around the fixed Earth. These
motions can be seen as real and self-sustained. If one could put the whole Universe in
accelerated motion around the Earth, the potentials (\ref{Evecpot}) and
(\ref{Ups}) would immediately be generated and would keep the Universe in that very
same state of motion \emph{ad infinitum}.

{\ack This work is supported by the Croatian Government under contract
number 119-0982930-1016.}

\end{document}